\documentstyle[12pt]{article}

\begin{document}
\title{Photon-added coherent states for exactly solvable Hamiltonian}
\author{{\bf M. Daoud} \\
\\ 
Abdus Salam International Centre of Theoretical Physics \\ Trieste, Italy\\
\\and\\
University Ibn Zohr, LPMC , Department of physics\\  Agadir, Morocco}
\maketitle

\begin{abstract}
 Photon-added Gazeau-Klauder and Klauder-Perelomov coherent states are investigated. Application to P\"oschl-Teller potentials is given.
\end{abstract}

\vfill
\newpage 

\section{Introduction}
Coherent states play an important role in quantum physics. In the recent years there are much interest in their applications and generalizations [1-3]. The original coherent states based on the Weyl-Heisenberg group has been extended to a number of Lie groups with square integrable representations. Recently, coherent states for exactly solvable quantum mechanical systems were introduced as eigenstates of the lowering operator. This definition leads to so-called Gazeau-Klauder coherent states [4-15]. Following the group theoretic method of Perelomov, another class of coherent states (Klauder-Perelomov coherent states) were considered [9,11,15]. To construct such coherent states, one deals with lowering and raising operators of the considered quantum system. In this respect, the one dimensionnal supersymmetric quantum mechanics (SSQM)[16] provides an algebraic tool to define creation and annihilation operators for an exact solvable potential. First, let us consider a Hamiltonian $H$ admitting a non degenerate discrete infinite energy spectrum. The energy levels of this system are assumed in increasing order
\begin{equation}
E_0 = 0 < E_1 <...< E_{n-1} < E_{n} <...
\end{equation}
It is well known that, for a sovable system, one can factorize the Hamiltonian as
\begin{equation}
H = A^+A^-
\end{equation}
The supersymmetric partners $H \equiv H_-$ and $H_+ =A^-A^+ $ have the same energy spectra, but different eigenstates. The Schr\"odinger equations for $H_{\pm}$  
\begin{equation}
H_{\pm}\vert \psi_n^{\pm}\rangle = E_n^{\pm}\vert \psi_n^{\pm}\rangle,
\end{equation}
implie that $A^{\pm} \vert \psi_n^{\pm}\rangle$ are eigenstates of $H_{\mp}$ with eigenvalues $E_n^{\pm}$ and $E_n^+ = E_{n+1}^- $. The $E_n^-$'s stand for the energy levels of $H$. The hamiltonians $H_{\pm}$ are isospectral and one gets
\begin{equation}
A^- \vert \psi_{n+1}^- \rangle = \sqrt {E_{n+1}^-} e^{i\alpha (E_{n+1}^-- E_{n}^-)} \vert \psi_{n}^+ \rangle
\end{equation}
\begin{equation}
A^+ \vert \psi_{n}^+ \rangle = \sqrt {E_{n}^+} e^{-i\alpha (E_{n}^+- E_{n-1}^+)} \vert \psi_{n+1}^- \rangle
\end{equation}
where $\alpha \in {\bf R}$. The operators $A^+$ and $A^-$ connect the states $\vert \psi_{n}^- \rangle$ and $\vert \psi_{n}^+ \rangle$ and they are not the ladder operators of the Hamiltonian $H_-$. To define the creation and annihilation operators for $H$, we consider the unitary transformation
\begin{equation}
U = \sum_{n,m = 0}^{\infty} U_{nm} \vert \psi_{n}^- \rangle \langle \psi_{m}^- \vert
\end{equation}
where $ U_{nm} = \langle \psi_{n}^- \vert  \psi_{m}^+ \rangle $. Note that $U^+ U = 1$ and $\vert  \psi_{n}^+ \rangle = U \vert  \psi_{n}^- \rangle$. Using the transformation $U$, define a new pair of operators
\begin{equation}
a^+ = A^+ U {\hskip 2cm},{\hskip 2cm} a^- = U^+ A^- 
\end{equation}
Combining (4), (5) and (6), we get the actions of $a^+$ and $a^-$ on the eigenstates $\{ \vert \psi_n^-\rangle \equiv \vert \psi_n\rangle\}$ as
\begin{equation}
a^- \vert \psi_{n} \rangle = \sqrt {E_{n}} e^{i\alpha (E_{n}- E_{n-1})} \vert \psi_{n-1} \rangle
\end{equation}
\begin{equation}
a^+ \vert \psi_{n} \rangle = \sqrt {E_{n+1}} e^{-i\alpha (E_{n+1}- E_{n})} \vert \psi_{n+1} \rangle
\end{equation}
The set of operators $\{ a^+ , a^-\}$ are the creation and annihilation operators of $H = a^+ a^- = A^+A^-$. They satisfy the following commutation relations
\begin{equation}
[ a^- , a^+ ] = a_0
\end{equation} 
where the operator $a_0$ is given by
\begin{equation}
a_0  \vert \psi_{n} \rangle  = ( E_{n+1} - E_n )  \vert \psi_{n} \rangle 
\end{equation}
The algebra generated by $a^+$ , $a^-$ and $a_0$ ( obtained here using supersymmetric techniques is similar to one derived in [6]) provides a mathematical tool leading to coherent states for an arbitrary quantum potential. This algebra permits a construction an important family of states obtained by adding photons (or more precisely excitations) to conventionnal coherent states. These states are called Photon-added coherent states. They were introduced by Agarwal and Tara [17,18] and extensively studied in [19,20]. Recently, Popov [21] purposed Photon-added coherent states of the pseudoharmonic oscillator by adding $k$ photons to  states; which are eigenvectors of the annihilation operator ( Gazeau-Klauder coherent states in the terminology of this work). The main task of this work is to give a general method to construct Photon-added coherent states by adding photons to Gazeau-Klauder and Klauder-Perelomov coherent states. The term Klauder-Perelomov do not should be confused with one previously used in [22] in another context. Note also that the term "Photon-added" is used in the sense discussed in [21]. Indeed, the commutation relation (10) is not canonical (except for harmonic oscillator) and the lowering/raising operators are not bosons (particularly photons) destruction/creation operators. Note that the word "Photon" is used, in this paper, instead "quanta" or "excitation". As illustration, we consider the case of a free particle embedded in a potential of P\"oschl-Teller type.  
\section{\bf Photon-added coherent states : General case}
In this section, we present a general method to construct Photon-added coherent states of Gazeau-Klauder and Klauder-Perelomov ones. For each kind, the kernel (scalar product of two states) and the measure in respect which the partition of unity is satified are given. Finally, we show their temporal stability.
\subsection{\bf Photon-added Gazeau-Klauder coherent states}
The Gazeau-Klauder coherent states are defined as eigenstates of lowering operator $a^-$
\begin{equation}
a^- \vert z , \alpha \rangle = z \vert z , \alpha \rangle
\end{equation}
where $z$ is an arbitrary complex number. The solutions of the eigenvalue equation (12) are given by
\begin{equation}
\vert z , \alpha \rangle = {\cal N}_0(\vert z \vert)^{-1/2}\sum_{n=0}^{\infty}\frac{z^n}{\sqrt{E_0(n)}}e^{-i\alpha E_n}\vert \psi_n\rangle
\end{equation}
where
\begin{equation}
{\cal N}_0(\vert z \vert) = \sum_{n=0}^{\infty} \frac{\vert z \vert^{2n}}{E_0(n)}
\end{equation}
stands for normalization constant and the energy function is defined by $E_0(n) = E_n E_{n-1} ... E_1$ for $n\ne 0$ and $1$ for $n=0$. \\
Photon-added coherent states are defined by repeated application of the creation operator $a^+$ to the Gazeau-Klauder coherent states
\begin{equation}
\vert z , \alpha ,k \rangle = {\cal N}_k(\vert z \vert)^{-1/2} (a^+)^k \vert z , \alpha \rangle 
\end{equation}
where $k$ is a positive integer being the number of added excitation to coherent states and $ {\cal N}_k(\vert z \vert)$ is the normalization constant. Using (8), (9) and (13), one gets
\begin{equation}
\vert z , \alpha ,k \rangle = {\cal A}_k(\vert z \vert)^{-1/2}\sum_{n=0}^{\infty}\frac{z^n}{\sqrt{E_k(n)}}e^{-i\alpha E_{n+k}}\vert \psi_{n+k}\rangle
\end{equation}
where $E_k(n)$ is defined as
\begin{equation}
E_k(n) = \frac{E_0(n)^2}{E_0(n+k)}
\end{equation}
and
\begin{equation}
{\cal A}_k(\vert z \vert) = {\cal N}_k(\vert z \vert){\cal N}_0(\vert z \vert)
\end{equation}
From normalization condition of the states (16), we get
$
{\cal A}_k(\vert z \vert) = \sum_{n=0}^{\infty} \frac{\vert z \vert^{2n}}{E_k(n)}
$
and we have
$
{\cal N}_k(\vert z \vert)  = \frac{1}{{\cal N}_0(\vert z \vert) } \sum_{n=0}^{\infty} \frac{\vert z \vert^{2n}}{E_k(n)}
$.
Notice that for $k=0$, we recover the Gazeau-Klauder coherent states. From (16), we deduce the overlapping, between two Photon-added coherent states, as
\begin{equation}
\langle z',\alpha ', k \vert z, \alpha , k \rangle = {\cal A}_k(\vert z \vert)^{-1/2}{\cal A}_k(\vert z' \vert)^{-1/2}
\sum_{n=0}^{\infty} \frac{ \bar z'^{n} z^{n}}{E_k(n)} e^{-iE_{n+k}(\alpha - \alpha ')}
\end{equation}
Here again for $k=0$, we recover the overlapping between two Gazeau-Klauder coherent states (13). An important property of the Photon-added coherent states (16) is the identity resolution
\begin{equation}
\int d\mu(z,\bar z)\vert z, \alpha , k \rangle \langle z ,\alpha , k \vert = \sum_{n=0}^{\infty} \vert \psi_{n+k}\rangle \langle \psi_{n+k} \vert
\end{equation}
where $d\mu(z,\bar z)$ is the measure to be determined. From the latter equation we see that the first states $\vert \psi_n \rangle$ with $n = 0, 1,..., k-1$ are absent from the expansion in the right hand side. The relation (20) is not identity resolution, since the operator in the right hand side is not unity operator. These means that the sum  sould be identified with the identity operator in order to have the completion. Assuming the isotropy condition of the measure, i.e.depends only on $\vert z \vert$, we set
\begin{equation}
d\mu(z,\bar z) = {\cal A}_k(\vert z \vert) h_k( \vert z \vert ^2) \vert z \vert d \vert z \vert d\theta /\pi
\end{equation}
with $z = \vert z \vert e^{i\theta}$. Remark that the integral in (20) is over the disk of radius of convergence
\begin{equation}
{\cal R}_k = \lim_{n\to\infty} \root n \of{ E_k(n)}
\end{equation}
Substituting (16) in the equation (20), one can easily verify that the function $h_k(u = \vert z \vert ^2)$ satisfy the following equation
\begin{equation}
\int_0^{{\cal R}_k^2 } h_k(u) u^n du = E_k(n) 
\end{equation}
In the particular situation where ${\cal R}_k \to +\infty$, the function $ h_k(u)$ is the Mellin transform of $E_k(s-1)$
\begin{equation}
h_k(u) = \frac{1}{2\pi i}\int_{c-i\infty}^{c+i\infty}E_k (s-1)u^{-s}ds {\hskip 1cm} c\in {\bf R}
\end{equation}
It is clear that the Photon-added coherent states, of Gazeau-Klauder type, are temporally stable. Indeed, we have
\begin{equation}
e^{-itH}\vert z, \alpha , k \rangle = \vert z, \alpha + t , k \rangle.
\end{equation}
\subsection{Photon-added Klauder-Perelomov coherent states}
The so-called Klauder-Perelomov coherent states for an arbitrary quantum system are defined 
\begin{equation}
\vert Z , \alpha \rangle = \exp ( Z a^+ - \bar Z a^-) \vert \psi_0 \rangle
\end{equation}
as the action of the displacement operator on the ground state wavefunction $ \vert \psi_0 \rangle$. The product of this action is given by [9]
\begin{equation}
\vert Z , \alpha \rangle = \sum_{n=0}^{\infty} \frac{Z^n}{\sqrt{F_n(\vert Z \vert)}} e^{-i \alpha E_n}\vert\psi_n\rangle
\end{equation}
where
\begin{equation}
F_n(\vert Z \vert^2) = \frac{1}{E_0(n)}\bigg[\sum_{j=0}^{\infty}\frac{(-|Z|^2)^j}{(n+2j)!}\bigg(\sum_{i_1=1}^{n+1}E_{i_1}\sum_{i_2=1}^{i_1+1}E_{i_2}...\sum_{i_j=1}^{i_{j-1}+1}E_{i_j}\bigg)\bigg]^{-2}
\end{equation}
in terms of the energies of the quantum system under consideration. The range of validity of the last expression depends on the spectrum of the quantum system under consideration. As in the previous case,
the Photon-added Klauder-Perelomov coherent states are introduced as 
\begin{equation}
\vert Z , k , \alpha \rangle = {\cal N}_k(\vert Z \vert)^{-1/2} (a^+)^k \vert Z ,  \alpha \rangle 
\end{equation}
By successive application of the raising operator $a^+$ on the Klauder-Perelomov coherent states, it yields
\begin{equation}
\vert Z , k , \alpha \rangle = {\cal N}_k(\vert Z \vert)^{-1/2} \sum_{n=0}^{\infty} \frac{Z^n}{\sqrt{F_{n}^{k}(\vert Z \vert)}} e^{-i \alpha E_{n+k}}\vert\psi_{n+k}\rangle
\end{equation}
where
\begin{equation}
F_{n}^{k}(\vert Z \vert) = F_n(\vert Z \vert^2)\frac{E_0(n)}{E_0(n+k)}
\end{equation}
In the particular case $k=0$, we recover the states (27). The normalization constant $ {\cal N}_k(\vert Z \vert)$ is given by
\begin{equation}
{\cal N}_k(\vert Z \vert) = \sum_{n=0}^{\infty}\frac{|Z|^2}{F_{n}^{k}(\vert Z \vert)}
\end{equation}
The kernel $\langle Z' , k , \alpha '\vert Z , k, \alpha \rangle$ is 
\begin{equation}
\langle Z' , k , \alpha '\vert Z , k, \alpha \rangle =\bigg[ {\cal N}_k(\vert Z \vert){\cal N}_k(\vert Z' \vert)\bigg]^{-1/2}
\sum_{n=0}^{\infty}
 \frac{\bar Z'^n}{\sqrt{F_{n}^{k}(\vert Z' \vert)}} \frac{Z^n}{\sqrt{F_{n}^{k}(\vert Z \vert)}}e^{-iE_{n+k}(\alpha - \alpha ')}
\end{equation}
Let us now look for the overcompletion property of states (30)
\begin{equation}
\int d\mu (Z,\bar Z) \vert Z , k, \alpha \rangle \langle Z , k , \alpha \vert = \sum_{n=0}^{\infty} \psi_{n+k}\rangle\langle\psi_{n+k}\vert
\end{equation}
Setting $d\mu (Z,\bar Z) {\cal N}_k(\vert Z \vert)^{-1} = h_k(\vert Z \vert^2)\vert Z \vert d \vert Z \vert/ \pi$, the overcompletion is guaranated if one can find a function $h_k$ satisfying
\begin{equation}
\int_{0}^{{\cal R}_k} h_k(\vert Z \vert^2) d \vert Z \vert ^2 \frac{ \vert Z \vert ^{2n}}{ F_n^k(\vert Z \vert ^2)} = \frac{E_0(n+k)}{E_0(n)}
\end{equation}
where ${\cal R}_k$ is the radius of convergence of the integral. For instance, in the case of the standard harmonic oscillator ${\cal R}_k = \infty$. Finally, one can verify that the Photon-added states (30) are temporally stable.
\section{P\"oschl-Teller potentials}
Now, we consider a one dimensional quantum system embedded in the P\"oshl-Teller potential
\begin{equation}
V_{\kappa,\kappa'}(x)= \frac{1}{4a^2}\bigg[ \frac{\kappa (\kappa - 1)}{\sin ^2(x/2a)} - \frac{\kappa' (\kappa' - 1)}{\cos ^2(x/2a)}
\bigg] - \frac{(\kappa + \kappa')^2}{4a^2}
\end{equation}  
for $0 < x < \pi a$ and vanishing elsewhere. The eigenvalues of the Hamiltonian $ H = -\frac{d^2}{dx^2} + V_{\kappa,\kappa'}(x)$ are $E_n = n(n+ \kappa + \kappa')$. The P\"oschl-Teller Hamiltonain can be factorized as $H = A^+A^-$ in terms of the operators 
\begin{equation}
A^{\pm} = \mp \frac{d}{dx} + W(x)
\end{equation}
where
\begin{equation}
W(x) = \frac{-1}{2a} \bigg[\kappa \cot(\frac{x}{2a}) - \kappa '\tan(\frac{x}{2a})\bigg]
\end{equation}
is the superpotential. The eigenstates of $H \equiv H_-$ are given by
\begin{equation}
\psi_n^- (x) = [c_n^{\kappa , \kappa '}]^{-1/2} \bigg(\cos(\frac{x}{2a})\bigg)^{\kappa '} \bigg(\sin(\frac{x}{2a})\bigg)^{\kappa } P_n^{\kappa - 1/2 , \kappa '- 1/2} \bigg(\cos(\frac{x}{a})\bigg)
\end{equation}
with the normalization constant $c_n^{\kappa , \kappa '}$ defined by
\begin{equation}
c_n^{\kappa , \kappa '} = a \frac{\Gamma (n + \kappa + 1/2)\Gamma (n + \kappa' + 1/2)}{\Gamma (n + 1)\Gamma (n + \kappa + \kappa ') (2n + \kappa + \kappa ')}
\end{equation}
The $P_n^{\alpha , \beta}$'s stands for the Jacobi polynomials. The eigenstates of the supesymmetric partner $H_+ = A^-A^+$ are given by 
\begin{equation}
\psi_n^+ (x) = [c_n^{\kappa + 1 , \kappa '+ 1}]^{-1/2} \bigg(\cos(\frac{x}{2a})\bigg)^{\kappa ' + 1} \bigg(\sin(\frac{x}{2a})\bigg)^{\kappa + 1 } P_n^{\kappa + 1/2 , \kappa '+ 1/2} \bigg(\cos(\frac{x}{2a})\bigg)
\end{equation}
At this stage, one can evaluate the matrix elements of the unitary transformation (6) which permits the definition of the creation and annihilation operators. They are given by
\begin{equation}
U_{nm} = \langle \psi_n^- \vert \psi_n^+ \rangle = \int_0^{\pi a} \psi_n^+ (x)\psi_n^- (x) dx
\end{equation}
Replacing in (42) the eigenstates $  \psi_n^-(x)$ and $ \psi_m^+ (x)$ by their expressions (Eqs.(39) and (40)) and using the definition of the Jacobi polynomials, one obtain
\begin{eqnarray}
U_{nm} =\lefteqn{ a [c_n^{\kappa , \kappa '}c_m^{\kappa + 1 , \kappa '+ 1}]^{-1/2}\sum_{p=0}^{n}\sum_{p'=0}^{m}{n+\kappa - \frac{1}{2}\choose p}{n+\kappa' - \frac{1}{2}\choose n-p}}\nonumber\\
&& {m+\kappa + \frac{1}{2}\choose p'}{m +\kappa' + \frac{1}{2}\choose m-p'} B(\kappa + p +p'+1,n+m+\kappa'+1-p)
\end{eqnarray}
where $B(\mu ,\nu )$ is the well known Beta function. Now, one introduce the ladder operators, for P\"oschl-Teller potential,
that act on the eigenvectors $\{\vert \psi_n \rangle\}$ as
\begin{equation}
a^+ \vert \psi_n \rangle = \sqrt {(n+1)(n+\kappa+\kappa'+1)} e^{-i\alpha (2n+\kappa+\kappa'+1)} \vert \psi_{n+1}\rangle
\end{equation}
\begin{equation}
a^- \vert \psi_n \rangle = \sqrt {n(n+\kappa+\kappa')} e^{i\alpha (2n+\kappa+\kappa'-1)} \vert \psi_{n-1}\rangle
\end{equation}
The right hand side of (43) would be written in a compact form by making use some special functions and their properties.
\subsection{\bf Gazeau-Klauder case}
The Photon-added Gazeau-Klauder coherent states, in the P\"oschl-Teller case, are 
\begin{equation}
\vert z , \alpha , k \rangle = {\cal A}_k(|z|^2)^{-1/2} \sum_{n=0}^{\infty} \frac {z^n}{n!}\frac{\sqrt{(n+k)!(\lambda+k+1)_n}}{(\lambda + 1)_n} e^{-i\alpha E_{n+k}}\vert \psi_{n+k}\rangle
\end{equation}
where $\lambda = \kappa + \kappa'$ and $(.)_n$ denotes Pochhammer's symbol. For $k=0$, one has the Gazeau-Klauder coherent states

\begin{equation}
\vert z , \alpha , k=0 \rangle = {\cal A}_0(|z|^2)^{-1/2} \sum_{n=0}^{\infty} \frac {z^n}{ \sqrt{n!(\lambda + 1)_n}} e^{-i\alpha E_{n}}\vert \psi_{n}\rangle
\end{equation}
From the normalization condition $\langle z , k ,\alpha \vert z , k , \alpha \rangle = 1$, we obtain the normalization constant
\begin{equation}
{\cal A}_k (|z|^2) = \Gamma(k+1)   {\hskip 0.1cm}_2F_3 (k+1, \lambda + k +1; 1 , \lambda + 1 , \lambda + 1 ; |z|^2)
\end{equation}
which reduces, for $k=0$, to the normalization factor of states (47) 
\begin{equation}
{\cal A}_0(|z|^2) = {\hskip 0.1cm}_0F_1(1 + \lambda ; |z|^2 )
\end{equation}
The overlapping between Photon-added coherent states (46) takes the following form
\begin{eqnarray}
\lefteqn{\langle z' , k ,\alpha' \vert z , k , \alpha \rangle = {\cal A}_k(|z'|^2)^{-1/2}{\cal A}_k(|z|^2)^{-1/2}}\nonumber\\
&&\sum_{n=0}^{\infty} \frac{\bar z'^n z^n}{n!} \frac{\Gamma(k+1)(k+1)_n(\lambda + k + 1)_n}{(1)_n(\lambda + 1)_n(\lambda + 1)_n}
e^{-iE_{n+k}(\alpha - \alpha')}
\end{eqnarray}
For $\alpha = \alpha'$, The scalar product (50) takes the following compact form
\begin{equation}
\langle z' , k ,\alpha' \vert z , k , \alpha \rangle = {\cal A}_k(|z'|^2)^{-1/2}{\cal A}_k(|z|^2)^{-1/2}
{\hskip 0.1cm}_2F_3 (k+1, \lambda + k +1; 1 , \lambda + 1 , \lambda + 1 ; \bar z' z )
\end{equation}
The resolution to identity is given by
\begin{equation}
\int d\mu (\bar z , z )\vert z, k, \alpha \rangle \langle z, k, \alpha \vert = \sum_{n=0}^{\infty} \vert \psi_{n+k} \rangle \langle \psi_{n+k}\vert ,
\end{equation}
where the integration is carried out over the complex plane.
To obtain the measure $d\mu (\bar z , z )$, we define an isotropic function $h(r = |z|^2)$ ($z = z e^{i\theta}$) by
\begin{equation}
d\mu (\bar z , z ){\cal A}_k (|z|^2)^{-1} = h(r)rdrd\theta /\pi
\end{equation}
satisfying the following sum
\begin{equation}
\int_{0}^{+\infty} h(r) r^n dr = \frac{(n!)^2(\lambda + 1)_n^2}{(n+k)!(\lambda + k + 1)_n}
\end{equation}
The theory of the inverse Mellin transform [23] provides us with the weight function, satisfying the equation (54), in terms of a Meijer $G$-function: 
\begin{equation}
h(r) = \frac{\Gamma(\lambda + k + 1)}{\Gamma(\lambda + 1)^2}
G_{2,4}^{4,0}\bigg( r \bigg\vert {\hskip 0.2cm}_{0 , {\hskip 0.2cm} 0 , {\hskip 0.2cm}\lambda ,{\hskip 0.2cm} \lambda }^{k, \lambda + k}\bigg)
\end{equation}
Finally, we conclude that the Photon-added coherent states (47) are continously labeled by the complex variable $z$, resolve the identity and they are temporally stable.

\subsection{\bf Klauder-Perelomov case}
It now remains to consider the Photon-added coherent states of Klauder-Perelomov kind. They are given by
\begin{eqnarray}
\lefteqn{\vert \xi , k , \alpha \rangle = (1-|\xi|^2)^{\frac{\lambda+1}{2}}({\cal N}_k (|\xi|^2))^{-1/2}{}}
\nonumber\\
&&{}\sum_{n=0}^{\infty}\frac{\xi^n}{\sqrt{n!}}\bigg[\frac{\Gamma(n+k+1)}{\Gamma(n+1)}\bigg]^{1/2}\bigg[\frac{\Gamma(n+k+2\lambda+1)}{\Gamma(2\lambda+1)}\bigg]^{1/2}e^{-i\alpha E_{n+k}}
\vert \psi_{n+k}\rangle
\end{eqnarray}
Here, we have used the Klauder-Peremolov coherent states derived in a previous work [9]. The new variable $\xi$ is defined by $\xi = \frac {Z}{|Z|}\tanh |Z|$. The analytical representations of (56) are defined in the disk of unit radius and the normalization constant is given by
\begin{equation}
{\cal N}_k (|\xi|^2) = (1-|\xi|^2)^{\lambda+1}\frac{\Gamma (k+1)\Gamma(\lambda + 1 + k)}{\Gamma(\lambda + 1)}{\hskip 0.2cm}_2F_1 (\lambda + k + 1, k + 1 ; 1 ; |\xi|^2)
\end{equation}
The standard Klauder-Perelomov coherent states
\begin{equation}
\vert \xi , \alpha \rangle = (1-|\xi|^2)^{\frac{\lambda+1}{2}}\sum_{n=0}^{\infty}\xi^n \bigg[ \frac{\Gamma(n+\lambda+1)}{\Gamma(n+1)\Gamma(\lambda+1)}\bigg]^{1/2}e^{-i\alpha E_{n}}
\vert \psi_{n}\rangle
\end{equation}
are recovered by taking $k=0$. The computation of the kernel of states (56) gives
\begin{eqnarray}
\lefteqn{\langle \xi', k , \alpha' \vert \xi , k , \alpha \rangle = ({\cal N}_k (|\xi'|^2))^{-1/2}({\cal N}_k (|\xi|^2))^{-1/2}{}}
\nonumber\\
&&{}(1-|\xi'|^2)^{\frac{\lambda+1}{2}}(1-|\xi|^2)^{\frac{\lambda+1}{2}}\sum_{n=0}^{+\infty}\frac{\bar \xi'^{n}\xi ^n}{n!}\frac{\Gamma(n+k+1)}{\Gamma(n+1)}\frac{\Gamma(n+1+\lambda +k)}{\Gamma(\lambda + 1)}
\end{eqnarray}
The unity resolution relation, in the Klauder-Perelomov case, is written as follows 
\begin{equation}
\int d\mu (\bar \xi , \xi )\vert \xi, k, \alpha \rangle \langle \xi, k, \alpha \vert = \sum_{n=0}^{\infty} \vert \psi_{n+k} \rangle \langle \psi_{n+k}\vert
\end{equation}
Making the polar decomposition $\xi = r e^{i\theta}$ and the ansatz
\begin{equation}
d\mu(\xi ,\bar \xi)(1-|\xi|^2)^{\lambda + 1}({\cal N}_k (|\xi'|^2))^{-1} = \frac{\Gamma(\lambda + 1)^2}{\Gamma(k+1)\Gamma(\lambda + k + 1)} h(r)rdrd\theta /\pi , 
\end{equation}
it follows that the requierement such that the states $\vert \xi, k, \alpha \rangle$ form a complete (in fact an overcomplete) set is equivalent to the resolution of a power moment problem. Namely, determine a positive weight function such that
\begin{equation}
\int_{0}^{1} h(r) r^{n-1} dr = \frac{\Gamma(n + 1)^2}{\Gamma(n + k + 1)\Gamma(n + \lambda + k + 1)}. 
\end{equation}
The function $h(r)$ satisfying the last equation, in the unit disk, is [23]
\begin{equation}
h(r) = \frac{(1-r)^{\lambda+ 2k - 1}}{\Gamma(\lambda + 2k + 1)} {\hskip 0.2cm}_2F_1 (k , \lambda +k , \lambda + k ; 1 - r)
\end{equation}
Using the integral representation of the hypergeometric function $_2F_1$, one can verify that the function (63) gives $h(r) = \frac{(1-r)^{\lambda - 1}}{\Gamma(\lambda + 1)}$ for $k=0$ and we obtain the measure $d\mu(\xi ,\bar \xi) = \frac{1}{(1 - | \xi |^2)^2} |\xi |d|\xi |d\theta /\pi$ in respect which the Klauder-Perelomov coherent states (58) constitute an overcomplete set. 
\section{\bf Discussion and outlook}
In conclusion, we have given a construction of the Photon-added coherent states associated with exactly solvable Hamiltonians. A discussion this construction for Gazeau-Klauder and Klauder-Perelomov coherent states is done. The example of P\"oschl-Teller potential is studied. It is remarkable that the Photon-added coherent states possess all properties of the usual coherent states as continuity, overcompletion, non-orthogonality and temporal stability. These properties can exploited for a mathematical use. Indeed, Gazeau-Klauder Photon-added coherent states can be used to define star product to provide a deformation-quantization \`a la Moyal in the spirit of works [24,25]. In other hand, the Photon-added coherent states of Klauder-Perelomov type can give another geometric setting to investigate a quantization in the Berezin sense [26]. 
 {\vskip 1.0cm}
{\bf Acknowledgements}: 
The author would like to thank the Condensed matter Section of the Abdus Salam-ICTP for hospitality. He would like to thank also A. Hegazi for reading this paper. He is gratefull to the referee for critical comments and suggestions which helped to improve the clarity and conciseness of the presentation.\\   
\vfill\eject

\end{document}